\documentclass[aps,pra,twocolumn,amsmath,amssymb,groupedaddress,longbibliography]{revtex4-1}
\usepackage{graphicx}
\usepackage{dcolumn}

\begin{document}

\title{Topological quantum transport and spatial entanglement distribution via disordered bulk channel}

\author{Shi Hu$^{1,2}$}

\author{Yongguan Ke$^{1,3}$}

\author{Chaohong Lee$^{1,2,4}$}
\altaffiliation{Email: lichaoh2@mail.sysu.edu.cn}

\affiliation{$^{1}$Guangdong Provincial Key Laboratory of Quantum Metrology and Sensing $\&$ School of Physics and Astronomy, Sun Yat-Sen University (Zhuhai Campus), Zhuhai 519082, China}

\affiliation{$^{2}$State Key Laboratory of Optoelectronic Materials and Technologies, Sun Yat-Sen University (Guangzhou Campus), Guangzhou 510275, China}

\affiliation{$^{3}$Nonlinear Physics Centre, Research School of Physics, Australian National University, Canberra, Australian Capital Territory 2601, Australia}

\affiliation{$^{4}$Synergetic Innovation Center for Quantum Effects and Applications, Hunan Normal University, Changsha 410081, China}

\begin{abstract}
We show how to implement quantum transport, generate entangled state and achieve spatial entanglement distribution via topological Thouless pumping in one-dimensional disordered lattices.
We introduce the on-site disorders to suppress the high-order resonant tunneling (which cause wave-packet dispersion) and realize dispersionless Thouless pumping.
The interplay between the on-site disorders and the band topology enables robust unidirectional transport.
We also demonstrate how to prepare spatially entangled two-particle NOON state via Hong-Ou-Mandel interference assisted by the Thouless pumping.
The quantum entanglement can be well preserved during the distribution between spatially separated sites.
Our system paves a way to wide applications of topology in quantum information process.
\end{abstract}

\date{\today}
\maketitle

\section{Introduction}\label{Sec1}
Topological quantum pumping, a transport phenomenon utilizing topological properties of a modulated system, has attracted intensive attention in recent years.
There are two typical kinds of topological pumping, adiabatic pumping via topological edge modes~\cite{YEKrausPRL2012,MVerbinPRB2015,JKAsboth2016,NLang2017,FMeiPRA2018,JLTambascoSciAdv2018,PBorossPRB2019,SLonghiPRB2019} and Thouless pumping via topological bulk bands~\cite{DJThoulessPRB1983}.
In the adiabatic pumping, the edge state at one side will be adiabatically coupled to bulk states and then transferred to edge state at the other side~\cite{YEKrausPRL2012, MVerbinPRB2015, JKAsboth2016, NLang2017, FMeiPRA2018, JLTambascoSciAdv2018}.
Unlike non-quantized adiabatic pumping via edge channel, the Thouless pumping is a quantized transport directly connected to the topological invariants of bulk bands, Chern numbers, which guarantee robust protection against perturbations~\cite{QNiuJPA1984}.
To date, the Thouless pumping has been experimentally demonstrated via cold atoms in modulated optical lattices~\cite{MLohseNP2016,SNakajimaNP2016}, and can be employed to explore topological phases~\cite{DJThoulessPRB1983,LWangPRL2013,FMeiPRA2014,YKeLPR2016,YKePRA2017,SHuPRB2019,LLin2019}.
It is interesting to apply topological pumping in quantum information.

Quantum entanglement plays a pivotal role in quantum information process, quantum communication and quantum metrology~\cite{RHorodeckiRMP2009,JWPanRMP2012,LPezzeRMP2018}.
Topological protected entangled NOON state and its transport via edge states have been reported~\cite{MCRechtsmanOptica2016,ABRedondoScience2018,MWangNanophotonics2019}.
By using the adiabatic pumping via edge channels, the well-known Hong-Ou-Mandel (HOM) interference~\cite{CKHongPRL1987} of photon pair has been observed, and spatial entangled states have been generated~\cite{JLTambascoSciAdv2018}.
However, the adiabatic pumping needs to be extremely slow for long-distance transport where the gap between edge channel and the bulk band are extremely small.
Furthermore, the adiabatic pumping is only robust to the disorder preserving some specific symmetries~\cite{NLang2017,FMeiPRA2018,PBorossPRB2019}.
Recently, entanglement generation and robust transport are proposed via Thouless pumping of interacting bosons in a complex ring-lead system~\cite{THaugCP2019}, which is difficult to realize in the state-of-art experiments.
However, the conventional Thouless pumping of noninteracting particles always has wave-packet dispersion due to the high-order tunneling, which hinders the transport of Fock states~\cite{YKeLPR2016,SHuPRB2019,JTangpanitanonPRL2016}.
It is desirable to achieve robust quantum transport, entanglement generation, and distribution via Thouless pumping in a simple and realistic system.

In this paper, we propose how to achieve unidirectional transport, entanglement generation and entanglement distribution via the Thouless pumping in one dimensional modulated lattices with deliberately introduced on-site disorders.
On one hand, the Thouless pumping is robust against modest on-site disorders less than the energy gap, that is, the quantization of centre-of-mass (COM) shift during one pumping cycle is preserved.
On the other hand, the on-site disorders supress the high-order resonant tunneling and support dispersionless transport.
Both the uniformly distributed and normally distributed on-site disorders work well in the topological transport.
We not only propose how to realize topological transport of Fock states via the Thouless pumping of noninteracting particles, but also demonstrate the generation of spatial NOON state and entanglement distribution by combining the Thouless pumping and Hong-Ou-Mandel (HOM) interference.
Two distant and disentangled particles in different bands are first shifted to a unit cell via Thouless pumping, and turn into a spatial NOON state by a sudden quench following adiabatic pumping.
At last the NOON state are distributed in long distance via Thouless pumping again.
Compared with Ref.~\cite{JLTambascoSciAdv2018}, the modest on-site disorders benefit the realization of entangled state and its distribution instead of a obstacle.
Moreover, the long-distance transport is more accessible since the adiabatic condition is easier to be satisfied.

This paper is organized as follows.
In Sec.~\ref{Sec2}, we introduce our model, a noninteracting tight-binding Rice-Mele model.
In Sec.~\ref{Sec3}, we study how to transport Fock state via disordered bulk channel.
In Sec.~\ref{Sec4}, we illustrate the HOM interference and topological spatial entanglement distribution.
We also investigate their robustness against on-site disorders.
In Sec.~\ref{Sec5}, we give a summary and discussion of our results.

\section{Rice-Mele Model}\label{Sec2}
\begin{figure}[!htp]
\centering
\includegraphics[width=0.5\columnwidth]{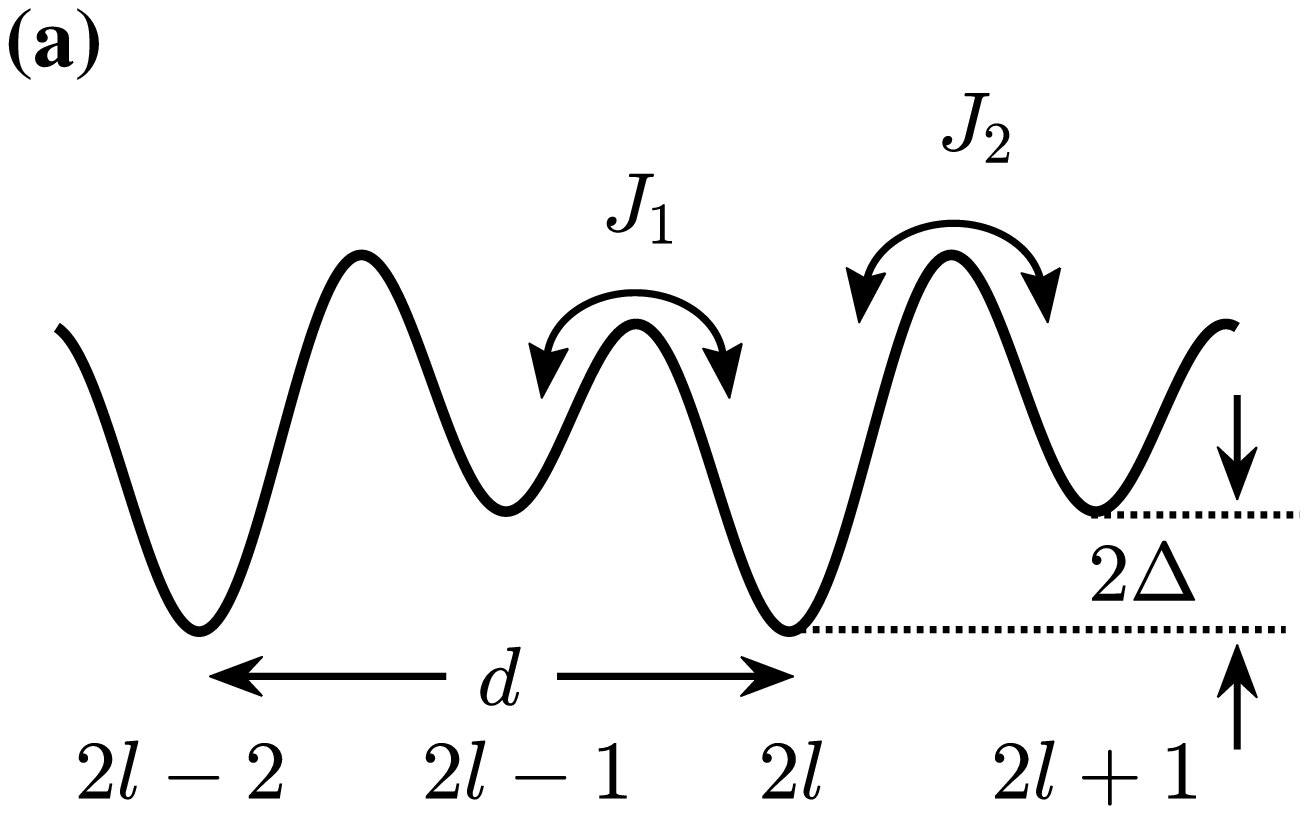}%
\includegraphics[width=0.5\columnwidth]{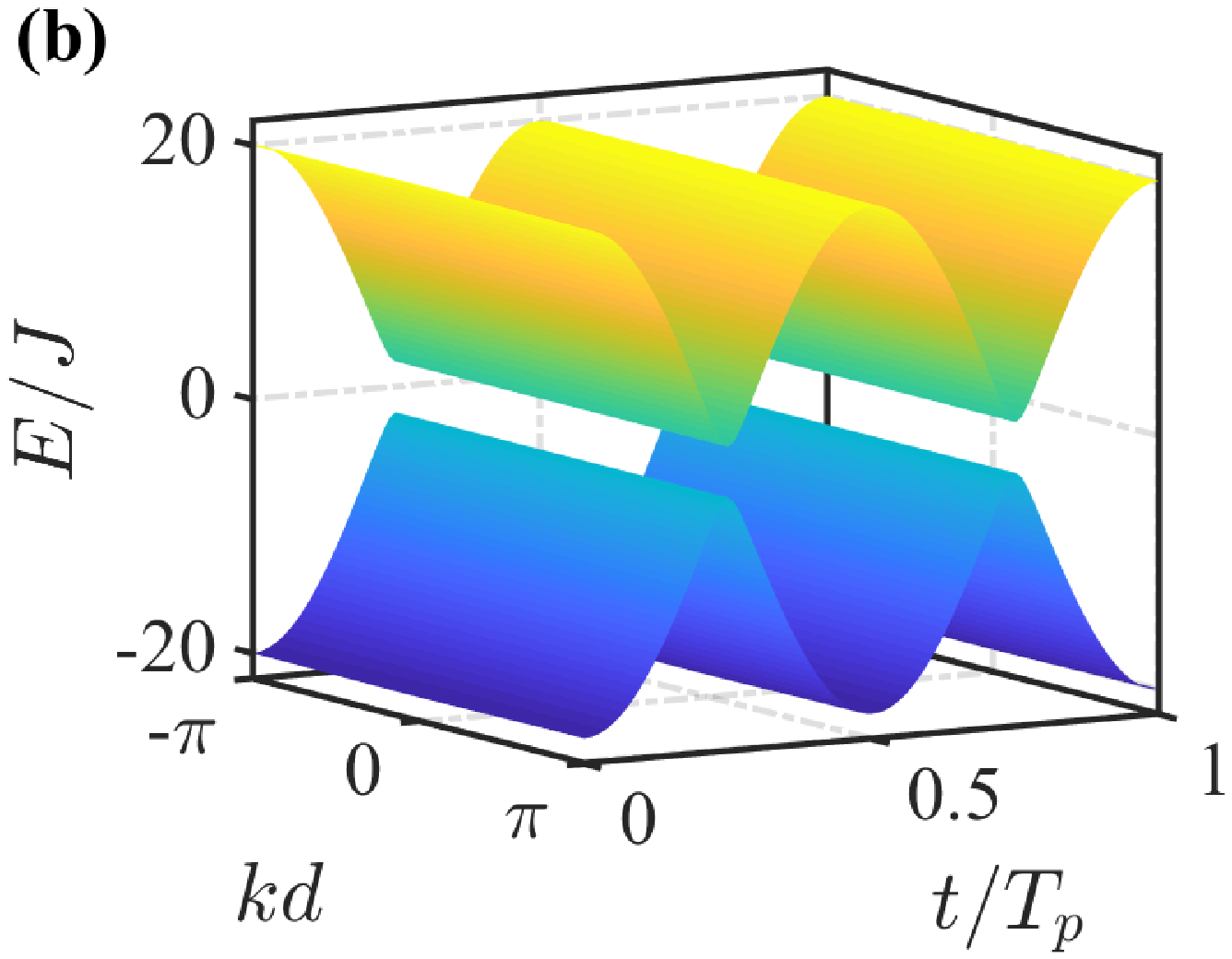}%
\caption{\label{RMSchematic}(color online).
(a) Schematic of the Rice-Mele model.
Here, $J_{1}$ and $J_{2}$ denote the tunnel couplings, $2\Delta$ is the energy offset between neighbouring sites, and \emph{d} is the length of unit cell.
(b) Energy bands structure in the $k-t$ Brillouin zone.
}
\end{figure}
As in the previous Thouless pumping experiments~\cite{MLohseNP2016,SNakajimaNP2016}, we consider the noninteracting tight-binding Rice-Mele model~\cite{MJRicePRL1982}
\begin{eqnarray}\label{Eq.Ham}
\hat{H}(t)&=&-\sum_{l=1}^{L}\Big[J_{1}(t)\hat{c}_{2l-1}^{\dag}\hat{c}_{2l}
+J_{2}(t)\hat{c}_{2l}^{\dag}\hat{c}_{2l+1}+{\rm H.c.}\Big]\cr\cr
&&+\sum_{l=1}^{L}\Delta(t)
\Big(\hat{c}_{2l-1}^{\dag}\hat{c}_{2l-1}-\hat{c}_{2l}^{\dag}\hat{c}_{2l}\Big),
\end{eqnarray}
with \emph{L} cells (the total sites is $2L$).
Here, $\hat{c}_{j}^{\dag}(\hat{c}_{j})$ is bosonic creation (annihilation) operator for the \emph{j}th site.
$J_{1}=J+\delta_{0}\sin\phi$ ($J_{2}=J-\delta_{0}\sin\phi$) denotes the intra (inter)-cell tunnelling amplitude, and $\Delta=\Delta_{0}\cos\phi$ is a staggered on-site energy offset, as shown in Fig.~\ref{RMSchematic} (a).
The time-dependent phase $\phi(t)$ is
adiabatically swept according to $\phi(t)=\omega t+\phi_{0}$ with ramping speed $\omega$ and initial modulated phase $\phi_{0}$.
The pumping period is $T_{P}=2\pi/\omega$.
In the following, we set $J=1$ and the other parameters $\Delta_{0}$, $\delta_{0}$, and $\omega$ are given in units of $J$.
We can derive the Hamiltonian $H(k,t)$ in the quasi-momentum space by a Fourier transformation.
Solving the eigenproblem $H(k,t)|\psi_{n}(k,t)\rangle=E_{k,n}(t)|\psi_{n}(k,t)\rangle$, we can obtain the energy band structure in the ($k-t$) space, as shown in Fig.~\ref{RMSchematic} (b).
The band gap is defined as
\begin{eqnarray}\label{Eq.Gap}
G(t)=\mathop{\min}\limits_{k}\big\{E_{2,k}(t)-E_{1,k}(t)\big\},
\end{eqnarray}
where $E_{1}$ ($E_{2}$ ) is the energy of the lower (upper) band.
The gap never closes during the whole pumping process, provided that $\delta_0$ and $\Delta_0$ are nonzero.

The topological features of Thouless pumping can be characterized by the Chern number,
\begin{eqnarray}\label{Eq.Chern number}
\nu_{n}=\frac{1}{2\pi}\int^{\pi/d}_{-\pi/d}dk
\int^{T_{P}}_{0}dt F_{n}(k,t),
\end{eqnarray}
which is defined within the $k-t$ Brillouin zone ($-\pi/d<k\leq\pi/d$, $0<t\leq T_{p}$)~\cite{DJThoulessPRB1983}.
Here,
$F_{n}(k,t)
=i(\langle\partial_{t}u_{n}|\partial_{k}u_{n}\rangle
-\langle\partial_{k}u_{n}|\partial_{t}u_{n}\rangle)$ is the Berry curvature and \emph{d} is the length of unit cell.
$|u_{n}(k,t)\rangle=\frac{1}{\sqrt{L}}
\sum_{j}u_{n,j}(k,t)c_{j}^{\dagger}|0\rangle$ is the periodic part of the Bloch state $|\psi_{n}(k,t)\rangle$.
Based on the single-particle polarization theory~\cite{RDKingPRB1993,DXiaoRMP2010}, the Chern number is connected to the shift of a single-particle Wannier state~\cite{GHWannierPR1937}, which is given as
\begin{eqnarray}
|W_{n}(R)\rangle_{1}
&=&\frac{1}{\sqrt{L}}\sum_{k}e^{-ikdR}|\psi_{n}(k)\rangle_{1}\cr\cr
&=&\frac{1}{L}\sum_{k,j}e^{-ikdR}e^{ikj}u_{m,j}(k)
c_{j}^{\dagger}|0\rangle_{1}.
\end{eqnarray}
where $R$ is the cell index, $n$ denotes the band index, and the ket subscript is a particle label.
The generalization for non-interacting particles is straightforward.
Without loss of generality, we consider two noninteracting bosons and the Wannier state is written as
\begin{eqnarray}\label{Eq.WannierState}
&&|W_{n,n'}(R,R')\rangle=\frac{1}{\sqrt{2(1+\delta_{n,n'}\delta_{R,R'})}}\cr\cr
&&\times\Big[|W_{n}(R)\rangle_{1}|W_{n'}(R')\rangle_{2}
+|W_{n'}(R')\rangle_{1}|W_{n}(R)\rangle_{2}\Big].
\end{eqnarray}
It is convenient to take the unique maximally localized Wannier states (MLWSs)~\cite{NMarzariPRB1997,NMarzariRMP2012}.
For a given Wannier state $|W_{n,n'}(R,R')\rangle$, the COM shift of the wave-packet in one pumping cycle is~\cite{RDKingPRB1993,DXiaoRMP2010}
\begin{eqnarray}\label{Eq.Shift}
\Delta P=\langle \hat{X}(T_{p})\rangle
-\langle \hat{X}(0)\rangle=d(\nu_{n}+\nu_{n'})/2,
\end{eqnarray}
where $\langle \hat{X}(t)\rangle=\langle W_{n,n'}(R,R',t)|\hat{X}|W_{n,n'}(R,R',t)\rangle$ is the COM position at the time \emph{t} and $|W_{n,n'}(R,R',t)\rangle$ is the instantaneous Wannier state.
The position operator is defined as $\hat{X}=N^{-1}\sum_{j=1}^{2L}j\hat{n}_{j}$ with $\hat{n}_{j}=\hat{c}_{j}^{\dag}\hat{c}_{j}$ and \emph{N} is the total particle number.
With this definition, we have $d=2$.

\section{Topological Quantum Transport}\label{Sec3}
\begin{figure}[!htp]
\centering
\includegraphics[width=1\columnwidth]{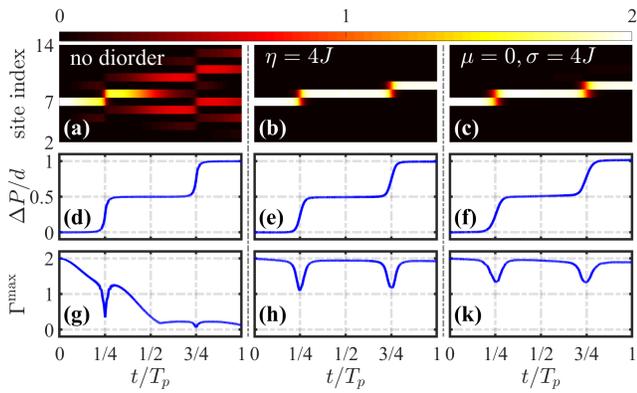}
\caption{\label{FockState}(color online).
Fock state transport with zero disorder (left), uniformly distributed disorder (middle), and normally distributed disorder (right), respectively.
The uniformly disorder amplitude $\eta=4J$ and the normally distributed disorder with mean parameter $\mu=0$ and standard variance $\sigma=4J$.
(a), (b), (c) The density distribution $\langle\hat{n}_{j}\rangle$, (d), (e), (f) the COM shift $\Delta P/d$, and (g), (h), (k) the maximum of diagonal two-particle correlation $\Gamma^{{\rm max}}$.
The two-particle Fock state is initialized at the site $2l-1=7$ and
the parameters are set as $2L=18$, $\Delta_{0}=20J$, $\delta_{0}=J$, $\omega=0.08J$, and $\phi_{0}=0$.
The results are averaged over 100 samples of on-site disorder.
}
\end{figure}
First, we discuss the topological transport of two-particle Fock state via disordered bulk channel.
At $t=0$ ($\Delta=\Delta_{0}, J_{1}=J_{2}=J, \phi_{0}=0$), the on-site energy offset is the largest and the MLWS of the lower (upper) band localized in \emph{l}th cell is residing only on the right (left) site.
For simplicity, we take the following denotation
\begin{eqnarray}
\frac{c_{j}^{\dagger}|0\rangle_{1}\otimes c_{j'}^{\dagger}|0\rangle_{2}+c_{j'}^{\dagger}|0\rangle_{1}\otimes c_{j}^{\dagger}|0\rangle_{2}}{\sqrt{2(1+\delta_{j,j'})}}= |j,j'\rangle.
\end{eqnarray}
We begin with a two-particle Wannier state
$c_{2l-1}^{\dagger}|0\rangle_{1}\otimes c_{2l-1}^{\dagger}|0\rangle_{2}=|2l-1,2l-1\rangle$, a Fock state with two particles in a site, and it fills the upper band.

We now evolve the system under the time-dependent Hamiltonian.
The transport is governed by the Chern number of the upper band ($\nu_{2}=+1$).
In Fig.~\ref{FockState}(a), we plot the density $\langle\hat{n}_{j}\rangle$ as a function of evolution time \emph{t}.
The two-particle Fock state is initially at the site $2l-1=7$
of an array with total size $2L=18$.
The parameters of the Hamiltonian are $\Delta_{0}=20J$, $\delta_{0}=J$, $\omega=0.08J$, and $\phi_{0}=0$.
The topological feature of the pumping is robust against the dispersion and the COM shift $\Delta P/d$ shows a clear step motion with $\nu_{2}=+1$ [Fig.~\ref{FockState} (d)].
However, due to the high-order resonant tunneling, each individual particle will be transported to several sites with the same potential during the pumping process.
The final state is dispersed and no longer a Fock state localized at certain sites [Fig.~\ref{FockState} (a)].
To explore the correlation feature, we calculate the two-particle correlation
\begin{eqnarray}
\Gamma_{q,r}=\langle\psi|\hat{c}_{q}^{\dag}\hat{c}_{r}^{\dag}
\hat{c}_{r}\hat{c}_{q}|\psi\rangle,
\end{eqnarray}
and the maximal value of diagonal correlation
\begin{equation}
\Gamma^{{\rm max}}=\mathop{\max}\limits_{q}\{\Gamma_{q,q}\}.
\end{equation}
For the initial two-particle Fock state $|2l-1,2l-1 \rangle$, the maximum of diagonal correlation is $\Gamma^{{\rm max}}=2$.
It decreases dramatically during the Thouless pumping and tends to 0 at the end of a pumping cycle [Fig.~\ref{FockState} (g)].

To suppress the dispersion caused by the high-order resonant tunneling, we separately consider two different kinds of on-site disorder, i.e., the uniformly distributed disorder and normally distributed disorder.
The uniformly distributed disorder is defined as
\begin{eqnarray}
\hat{H}_{\eta}=\eta\sum_{j}r_{j}\hat{n}_{j},
\end{eqnarray}
where $r_{j}\in[-1,1]$ is a uniformly distributed random number for each site and $\eta$ is the disorder amplitude.
While the normally distributed disorder is defined as
\begin{eqnarray}
\hat{H}_{\mu,\sigma}=\sum_{j}r_{j}^{\mu,\sigma}\hat{n}_{j},
\end{eqnarray}
where $r_{j}^{\mu,\sigma}$ is a random number from the normally distribution with mean parameter $\mu$ and standard variance $\sigma$.
In Fig.~\ref{FockState}(b), we plot the density $\langle\hat{n}_{j}\rangle$ with uniformly distributed disorder and the amplitude $\eta=4J$.
The disordered strength is large enough that the effective high-order tunneling becomes off-resonant and can be neglected, and it is small enough that the energy band gap maintains open in the whole pumping process.
In contrast to the ordered case, there is a step-like motion in the evolution of density distribution.
The final density is localized in the 9th site, corresponding to a transport of a unit cell length ($\nu_{2}=+1$).
The quantized COM shift $\Delta P/d$ still have a clear step-like motion under on-site disorder, see Fig.~\ref{FockState} (e).
After a pumping cycle, the correlation $\Gamma^{{\rm max}}$ is still close to 2 and it means that the final state is almost a Fock state $|\psi_{{\rm FOCK}}\rangle=|9,9\rangle$ [Fig.~\ref{FockState} (h)].
The results with normally distributed disorder are similar and we plot the density $\langle\hat{n}_{j}\rangle$, COM shift $\Delta P/d$, and two-particle correlation $\Gamma^{{\rm max}}$ in Fig.~\ref{FockState} (c), (f), and (k) respectively.
The parameters of the disorder are $\mu=0$ and $\sigma=4J$.

\begin{figure}[!htp]
\centering
\includegraphics[width=\columnwidth]{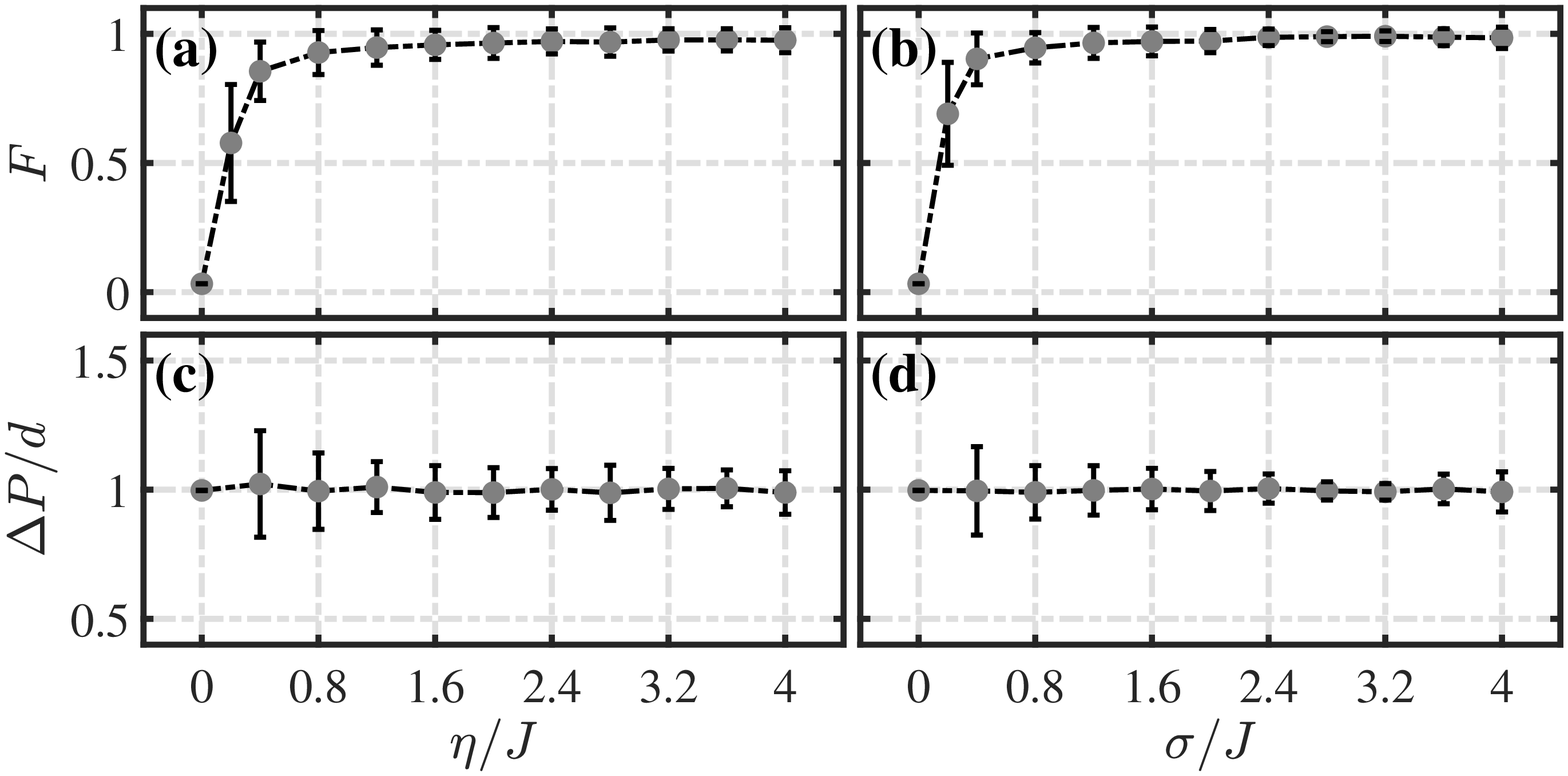}%
\caption{\label{FockDisorder}(color online).
(a),(b) Fidelity $F$ with uniformly and normally distributed disorder respectively.
(c),(d) COM shift $\Delta P/d$ with uniformly and normally distributed disorder respectively.
We set $\mu=0$ for the normally distributed disorder.
The parameters and the initial state are the same as the corresponding ones in Fig.~\ref{FockState}.
Each point is the average of 100 data sets and the
error bar shows the standard deviation.
}
\end{figure}

To explore how the uniformly and normally distributed disorder affect the topological transport, we numerically calculate the fidelity $F=|\langle\psi_{{\rm FOCK}}|\psi(T_{p})\rangle|$ as a function of the disorder amplitude $\eta$ and standard variance $\sigma$ by averaging a disordered ensemble of 100 samples, see Fig.~\ref{FockDisorder} (a) and (b),  respectively.
Here, $|\psi(T_{p})\rangle$ is the state after a pumping cycle.
The fidelity increases from $F=0$ to $F\approx1$ with $\eta$ and $\sigma$.
A wide plateau at $F\approx1$ appears for $(\eta,\sigma)\geq2J$, where
the disorder amplitude is large enough to suppress the high-order resonant tunneling.
Moreover, the standard variance of the fidelity decreases with $\eta$ and $\sigma$.
It tends to be stable for uniformly distributed disorder when $\eta\geq 2J$ and reaches a minimum around $\sigma=3J$ for normally distributed disorder.
The quantized COM shift $\Delta P/d$ is robust against disorder amplitude up to $(\eta,\sigma)=4J$ as shown in Fig.~\ref{FockDisorder} (c) and (d).
The variation of standard variance for the COM shift is similar to that of the fidelity.
The interplay between on-site disorder and the adiabatic pumping enables robust transport of Fock states with few particles in a site.
%
\section{Hong-Ou-Mandel Interference and Topological Spatial Entanglement Distribution}\label{Sec4}
\begin{figure}[!htp]
\centering
\includegraphics[width=\columnwidth]{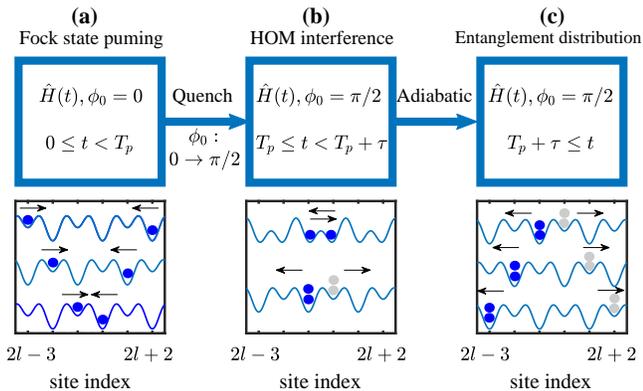}%
\caption{\label{HOMSchematic}(color online).
Schematic representation of the entire time evolution
of the system.
We separate the time evolution of the system into three
stages, i.e., (a) Fock state pumping, (b) HOM interference, and (c) entanglement distribution.
}
\end{figure}
In this section, we focus on the two-particle HOM interference and the spatial entanglement distribution via topological Thouless pumping.
We separate the time evolution of the system into three stages, i.e., Fock state pumping ($0 \le t\le nT_p$), HOM interference ($nT_p<t\leq nT_p+\tau$), and spatial entanglement distribution ($nT_p+\tau\le t\le 2nT_p + \tau$).
For simplicity, Fig.~\ref{HOMSchematic} shows schematic diagram of the pumping and interference processes for $n=1$.
As illustrated in Sec.~\ref{Sec3}, we introduce on-site disorder to suppress the wave-packet dispersion during the pumping.
To perform fast and efficient state pumping, HOM interference, and entanglement distribution, we change the phase $\phi(t)$ with evolution time according to the instantaneous energy gap, i.e., $\phi(t)=\epsilon\int_{0}^{t}G(t)dt+\phi_{0}$.
Here, since there is no interaction, $G(t)$ is the energy gap in the single-particle case.
The corresponding pumping period $T_{P}$ satisfies $\epsilon\int_{0}^{T_{P}}G(t)dt=2\pi$.

Without loss of generality, we begin the evolution at $t=0$ with $\phi_{0}=0$.
The two identical bosonic particles respectively localize in $(2l-3)$th and $(2l+2)$th sites, i.e., the initial state is $|2l-3,2l+2\rangle$.
At time $t=0$ the particle localized in $(2l-3)$th site uniformly fills the upper band ($\nu_{2}=+1$) and the other one uniformly fills the lower band ($\nu_{1}=-1$).
Then, we adiabatically tune the phase $\phi(t)$ in a nonlinear way and the two particles will move towards each other.
After a pumping cycle at $t=T_p$, the two particles are in the nearest neighboring sites, $(2l-1)$th and $2l$th sites, i.e. $|2l-1,2l \rangle$.
The on-site energy offset becomes the largest again [Fig.~\ref{HOMSchematic} (a)].

At $t=T_p$, we suddenly quench the system by changing the phase $\phi_{0}$ from 0 to $\pi/2$  and the on-site energy offset is vanished [Fig.~\ref{HOMSchematic} (b)].
Then, we adiabatically tune the phase $\phi(t)$ according to the instantaneous energy gap again.
After a duration time $t=\tau$ which satisfies $\epsilon\int_{0}^{\tau}G(t)dt=\pi/2$, the state evolves into a two-particle NOON state $(|2l-1,2l-1\rangle-|2l,2l\rangle)/\sqrt{2}$, see Fig.~\ref{HOMSchematic} (b).
The spatial NOON state is a maximally entangled state between the neighbor sites in a cell, due to the well-known HOM interference of two bosonic particle.
The quantum quench is critical to generate the NOON state and we will discuss in detail latter.
Then we adiabatical sweeping the phase $\phi(t)$ again according to the instantaneous energy gap.
We can distribute the entanglement between spatially separated sites via topological Thouless pumping.
After another pumping cycle, the final state at $t=\tau+2 T_{p}$ is given as $\big(|2l-3,2l-3\rangle-|2l+2,2l+2\rangle\big)/\sqrt{2}$, and the two-particle entanglement is separated by two unit cell [Fig.~\ref{HOMSchematic} (c)].

We now turn to discussion the HOM interference in detail.
To understand the process better we begin with a single particle at $(2l-1)$th site.
After the quantum quench ($\Delta=0, \phi_{0}=\pi/2$), the state $|2l-1\rangle$ can be decomposed as the superposition of a symmetric and antisymmetric single-particle state, $|2l-1\rangle=\frac{1}{\sqrt{2}}(|\psi\rangle_{S}+|\psi\rangle_{A})$, where $|\psi\rangle_{S}=\frac{1}{\sqrt{2}}(|2l-1\rangle+|2l\rangle)$ and $|\psi\rangle_{A}=\frac{1}{\sqrt{2}}(|2l-1\rangle-|2l\rangle)$.
Similarly, $|2l\rangle=\frac{1}{\sqrt{2}}(|\psi\rangle_{S}-|\psi\rangle_{A})$.
The symmetric (antisymmetric) single-particle state fills the lower (upper) band~\cite{MLohseNP2016}.
After a duration time $t=\tau$, the symmetric and antisymmetric states are respectively transferred to $|2l-1\rangle$ and $|2l\rangle$.
The pumping realizes a balanced beam splitter according to the transformation
$|2l-1\rangle\rightarrow
\frac{1}{\sqrt{2}}(e^{i\varphi_-}|2l-1\rangle+e^{i\varphi_+}|2l\rangle)$,
and $|2l\rangle\rightarrow
\frac{1}{\sqrt{2}}(e^{i\varphi_-}|2l-1\rangle-e^{i\varphi_+}|2l\rangle)$.
Here, $\varphi_{\pm}$ are the dynamical phases for the upper and lower bands respectively.
As a result, the two identical particles in the state $|2l-1,2l\rangle$ will evolves to
$\frac{1}{\sqrt{2}}(|2l-1,2l-1\rangle-e^{i2(\varphi_+-\varphi_-)}|2l,2l\rangle)$.
The dynamical phases accumulated in the process do not affect the performance of the HOM interference.
The HOM interference results from both the sudden quench and the topological pumping.
The sudden quench splits the initial statse as the equal superposition of symmetric and antisymmetric states which play the role of two interference pathes.
Without quantum quench, the two particles will always fill the lower and upper band respectively.
The final state is still $|2l-1,2l\rangle$ after sweeping the phase $\phi(t)$ from $\pi/2$ to $\pi$ adiabatically.
While the topological pumping due to the band topology plays the role of dynamical beamsplitter.
In contrast to the conventional beamsplitter where the Hamiltonian is static, we do not need to precisely control the duration time to realize HOM interference.
It is worth noting that there is a topological resonant tunneling for interacting bosons with suitable interaction strength and the HOM will disappear~\cite{YKeLPR2016}.

\begin{figure}[!htp]
\centering
\includegraphics[width=1\columnwidth]{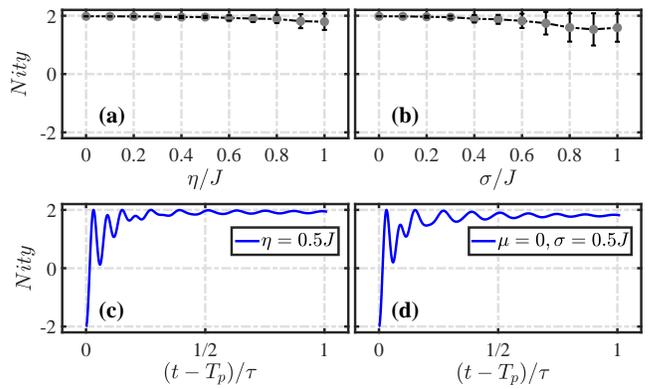}%
\caption{\label{HOMDisorder}(color online).
(a), (b) `NOONity' ($Nity$) at the end of HOM interference with uniformly and normally distributed disorder respectively.
Each point is the average of 100 data sets and the
error bar shows the standard variance.
(c),(d) $Nity$ as a function of evolution time during HOM interference with disorder amplitude $\eta=0.5J$ and $\mu=0$, $\sigma=4J$ respectively.
The results are averaged over 100 samples.
The two-particle initial state is $|9,10\rangle$ and
the parameters are set as $2L=18$, $\Delta_{0}=20J$, $\delta_{0}=J$, $\epsilon=0.03J$.
}
\end{figure}

To characterize the path entanglement of the generated NOON state, we calculate the `NOONity'  ($Nity$)~\cite{MCRechtsmanOptica2016}
\begin{eqnarray}
Nity=\sum_{q,r}\Gamma_{q,q}\Gamma_{r,r}-\Gamma_{q,r}^{2}.
\end{eqnarray}
This quantity will be larger if the state is more similar to the NOON state.
$Nity=2$ for an ideal NOON state and $Nity=-2$ for $|2l-1,2l\rangle$.
At the end of HOM interference, we respectively calculate the $Nity$ as a function of the disorder amplitude $\eta$ [Fig.~\ref{HOMDisorder} (a)] and standard variance $\sigma$ with $\mu=0$ [Fig.~\ref{HOMDisorder} (b)].
Here we only consider the HOM interference process and begin at $t=T_p$ with the input state $|9,10\rangle$.
For each point, the values are averaged over an ensemble of 100 samples.
Although an ideal HOM interference requires quenching to the lattices with vanished on-site energy offset, the on-site disorder has a tiny effect on the formation of NOON state.
The $Nity$ decreases slowly with disorder amplitude $\eta$ and the final $Nity\approx1.8$ when $\eta=J$.
The state is still a highly spatially entangled state with uniformly distributed disorder.
In contrast, the $Nity$ decreases more rapidly and it decrease to $Nity\approx1.4$ when $\sigma=1$ in the present of normally distributed disorder.
Moreover, the standard deviation of the $Nity$ also increases more rapidly than that with uniformly distributed disorder.

The HOM interference is not only robust against disorder, but also less sensitive to the duration time of the interference.
In Fig.~\ref{HOMDisorder} (c),(d), we also calculate $Nity$ as a function of evolution time $t$ during HOM interference with disorder amplitude $\eta=0.5J$ and $\mu=0$, $\sigma=4J$ respectively.
The $Nity=-2$ for the input state $|9,10\rangle$ and then it increases close to 2 at $t=\tau+T_p$.
Indeed we generate a NOON state by the HOM interference.
There is oscillation of $Nity$ in the first half duration and it is negligible at the end of the interference.
As illustrated in Fig.~\ref{HOMDisorder} (c),(d), there is no need to precisely control the interference time and a wide plateau at $Nity\approx2$ appear for $t-T_p\geq\tau/2$.

\begin{figure}[!htp]
\centering
\includegraphics[width=1\columnwidth]{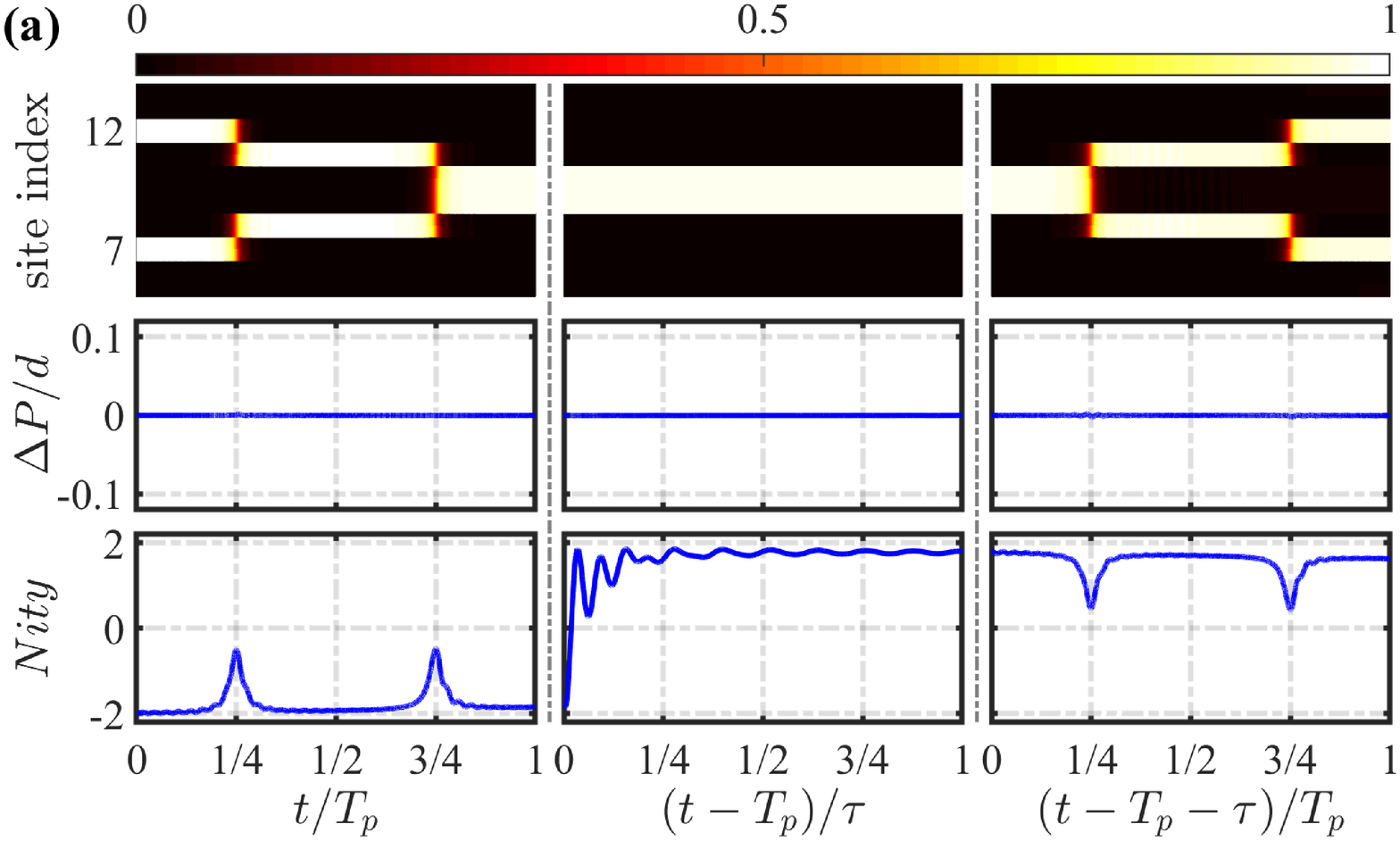}%
\hspace{0.05\columnwidth}%
\includegraphics[width=1\columnwidth]{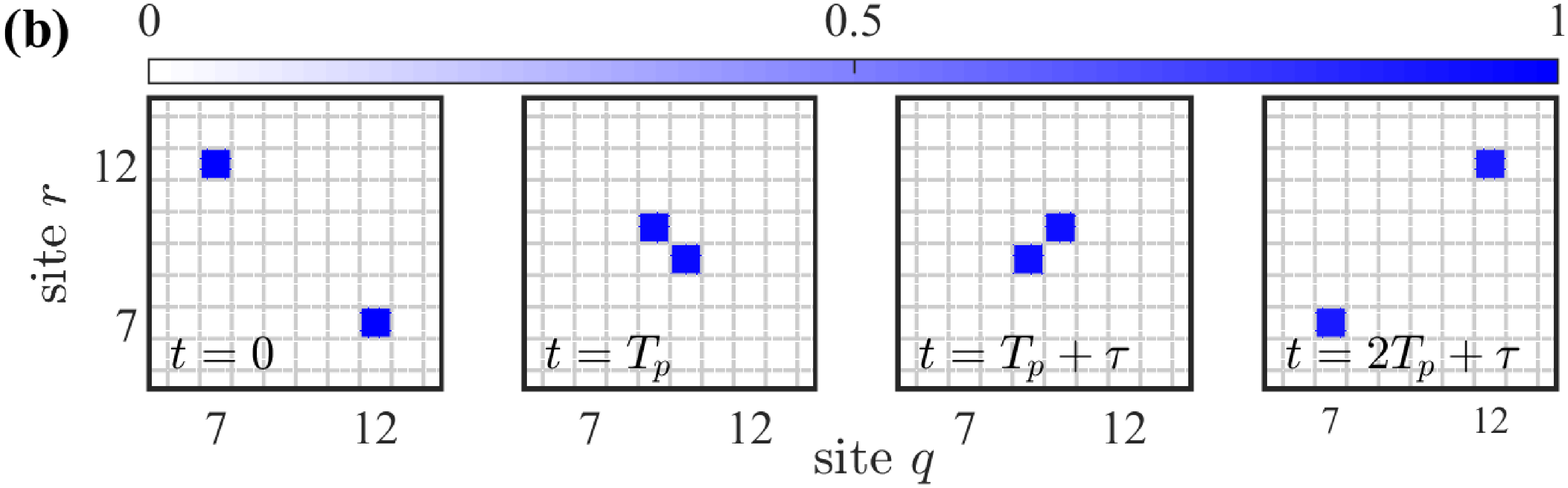}%
\caption{\label{HOMEffect}(color online).
Topological Thouless pumping with on-site disorder.
(a) Fock state pumping (left), HOM interference (middle), and spatial entanglement distribution (right).
We plot the density distribution $\langle\hat{n}_{j}\rangle$ (top), the COM shift $\Delta P/d$ (middle), and the two-particle $Nity$ (bottom) as function of evolution time.
(b) Two-particle correlation $\Gamma_{q,r}$ at certain evolution times.
The disorder amplitude $\eta=0.5J$.
The two-particle initial state is $|7,12\rangle$ and
the parameters are the same as those in Fig.~\ref{HOMDisorder}
The results are averaged over 100 samples of uniformly distributed on-site disorder.
}
\end{figure}

In Fig.~\ref{HOMEffect} (a), we plot the density distribution $\langle\hat{n}_{j}\rangle$ (top), the COM shift $\Delta P/d$ (middle), and the two-particle $Nity$ (bottom) as functions of evolution time $t$ during the entire process.
We only consider the uniformly distributed disorder and set the disordered amplitude $\eta=0.5J$ during the whole evolution.
The input state is an unentangled state, $|7,12\rangle$, which uniformly fills both of the two bands.
The parameters are set as $2L=18$, $\Delta_{0}=20J$, $\delta_{0}=J$, $\epsilon=0.03J$.
The density distribution of the input state and the final state are almost the same despite their huge difference in entanglement feature.
At initial time $t=0$, the two identical particles fill lower ($\nu_{1}=-1$) and upper ($\nu_{2}=+1$) bands, respectively.
Since the topological invariants for the upper and lower bands are opposite,  there is no COM shift during the pumping process in the first stage.
After the second stage, the state becomes entangled NOON state, which has the equal probability filling both lower and upper bands.
In conclusion there is no COM transport during the whole pumping process [middle row of Fig.~\ref{HOMEffect} (a)].
For the input state $|7,12\rangle$ the $Nity=-2$ and it always close to $-2$ in the stage of Fock state pumping.
Then it increase dramatically to $Nity\approx2$ in the stage of HOM interference.
At last, we realize the spatial entanglement distribution between the 7th and 12th sites via Thouless pumping.
The entanglement is well preserved and the final $Nity$ is still close to 2.
The peaks and dips of $Nity$ in the evolution appear around the time when the particles tunnel from one site to the other [bottom row of Fig.~\ref{HOMEffect} (a)].
We also plot the two-particle correlation $\Gamma_{q,r}$ at several typical evolution time at $0$, $T_p$, $T_p+\tau$, and $2T_p+\tau$, see Fig.~\ref{HOMEffect} (b).
The correlation function changes from uncorrelated anti-bunching to correlated bunching patterns, consistent with the \emph{Nity} evolution in Fig.~\ref{HOMEffect} (a).

\section{Summary and Discussion}\label{Sec5}

In summary, we have demonstrated that unidirectional quantum transport, spatial entanglement generation, and spatial entanglement distribution via topological Thouless pumping in disordered bulk channels.
The on-site disorders suppress the high-order resonant tunneling which causes wave-packet dispersion, and support dispersionless Thouless pumping.
As an adiabatic proposal, our spatial entanglement generation is not sensitive to the time duration of the interference.
The quantized, unidirectional, and dispersionless quantum transport and topological spatial entanglement distribution have wide potential applications in quantum information process and quantum metrology.
We also note that there has been great interests in exploring the interplay of disorder, topology, and quantum transport, such as the topological Anderson insulator~\cite{FEversRMP2008,EJMeierScience2018,SStutzerNature2018}.
Hence, it would be interesting to study topological quantum pumping in the topological Anderson insulator.
Moreover, our protocols are applicable for other topological models such as the Aubry-Andr\'{e}-Harper (AAH) model~\cite{PGHarperPPSA1955,SAubryAIPS1980}.

Lastly, we briefly discuss the experimental feasibility.
The Thouless pumping of noninteracting ultracold atoms in an optical superlattice has recently been demonstrated in experiments~\cite{MLohseNP2016,SNakajimaNP2016}.
The controllable on-site disorder can be easily introduced by an additional incommensurate lattice~\cite{GRoatiNature2008,MSchreiberScience2015,ALukinScience2019} or optical speckle~\cite{JBillyNature2008}.
Photonic lattices is another potential platform for testing our protocols.
It has been proposed that photonic waveguide could mimic AAH model and achieve adiabatic pumping~\cite{YEKrausPRL2012}.
In those photonic waveguides, disorder can be readily introduced by the state-of-art laser-written technique~\cite{SStutzerNature2018}.
Generation of NOON state in photonic waveguide systems may find wide applications in quantum communication and quantum information process.

\acknowledgements{This work is supported by the Key-Area Research and Development Program of GuangDong Province under Grant No.2019B030330001, the National Natural Science Foundation of China (NNSFC) under Grants No. 11874434 and No. 11574405, and the Science and Technology Program of Guangzhou (China) under Grant No. 201904020024. Y.K. was partially supported by the Office of China Postdoctoral Council under Grant No. 20180052 and the National Natural Science Foundation of China (NNSFC) under Grant No. 11904419.}

\end{document}